\documentclass[aps,prl,reprint,groupedaddress]{revtex4-2}
\usepackage{graphicx,hyperref,amsmath,xcolor}
\hypersetup{
       pdfauthor={Maximilian Patzauer, Katharina Krischer},
       pdftitle={
       Self-organized Multi-Frequency Clusters in an Oscillating Electrochemical System with Strong Nonlinear Coupling}
       pdfsubject={ },
       pdfkeywords={nonlinear dynamics, chimera, electrodissolution, pattern formation, multi-frequency cluster, coexistence patterns, oscillation, travelling waves, incoherence, amplitude death},
       breaklinks=true
}
\begin{document}
\title{
Self-organized Multi-Frequency Clusters in an Oscillating Electrochemical System with Strong Nonlinear Coupling}
\author{Maximilian Patzauer}
\email[]{maximilian.patzauer@tum.de}
\affiliation{
Nonequilibrium Chemical Physics, Department of Physics, Technical University of Munich, 85747 Garching, Germany
}
\author{Katharina Krischer}
\email[]{krischer@tum.de}
\affiliation{
Nonequilibrium Chemical Physics, Department of Physics, Technical University of Munich, 85747 Garching, Germany
}
\begin{abstract}
We study the spatio-temporal dynamics of the oscillatory photo-electrodissolution of n-type Si in a fluoride-containing electrolyte under anodic potentials using in-situ ellipsometric imaging.
When lowering the illumination intensity step-wise, we successively observe uniform oscillations, modulated amplitude clusters, and the coexistence of multi-frequency clusters i.e., regions with different frequencies, with a stationary domain.
We argue that the multi-frequency clusters emerge due to an adaptive, nonlinear, and nonlocal coupling, similar to those found in the context of neural dynamics.
\end{abstract}
\maketitle
Much of the dynamics of oscillating systems as diverse as neural activities \cite{Golomb2001}, electrical power grids \cite{Anvari2020}, multi-mode lasers \cite{Ludge2011}, and Josephson junction arrays \cite{Wiesenfeld1998} can be understood within the common framework of networks of coupled oscillators. 
These diverse applications render the study of coupled oscillators an important discipline of nonlinear dynamics. 
The overwhelming majority of previous studies consider the case of a linear coupling. 
Only recently, the more general case of nonlinear coupling has received increasing attention \cite{Rosenblum2007, Miethe2009, Temirbayev2012, Temirbayev2013, Schoenleber2014, Schmidt2014, Schmidt2015, Schmidt2015b,Komarov2015, Bolotov2018}.
It could be shown that this generalization can produce genuine nonlinear coupling features. 
For example, the case of global nonlinear coupling has been found to produce self-organized quasi-periodicity in ensembles of phase-oscillators \cite{Rosenblum2007}, complex chimera states composed of (nearly) synchronized regions of different mean frequencies and incoherent regions \cite{Bolotov2018}, or a variety of coexistence patterns, including again chimera states \cite{Schoenleber2014, Schmidt2014, Schmidt2015,Schmidt2015b}. 
Another representation of nonlinear coupling is realized in networks of Kuramoto oscillators that are designed to mimic adaptive neural networks \cite{Ren2007}.
In this type of model, recent studies predict the emergence of multi-frequency clusters even when the oscillators are identical \cite{Kasatkin2017,Berner2018,Berner2019}.
Until then, the existence of multi-frequency clusters was always linked to heterogeneous oscillatory systems with some distribution of the natural frequencies \cite{Osipov1997,Osipov1998a,Mikhailov2004,Menzel2010}. 

In this Letter, we report the emergence of self-organized multi-frequency clusters from a uniform oscillatory state during the photo-electrodissolution of an n-Si wafer when reducing the illumination intensity. 
Through the illumination, valence-band holes are created. 
Their movement parallel to the surface constitutes a nonlocal spatial coupling \cite{Patzauer2017}.
In addition, an external resistance in series with the electrode acts as a global synchronizing force on the dynamics \cite{Krischer2002}.
Hence, there are two dominant types of coupling, a global synchronizing coupling and a long-range coupling through diffusion and migration of valence-band holes. Below we will argue that their interaction creates a nonlinear coupling that, in turn, promotes the formation of multi-frequency clusters.

The oscillatory photo-electrodissolution of Si in fluoride-containing electrolytes involves the electrochemical oxidation of Si to SiO$_2$ according to
\begin{equation}
	\mathrm{Si+2H_{2}O}+\lambda_{\mathrm{VB}}h^{+}\rightarrow \mathrm{SiO_{2}+4H^{+}}+\left(4-\lambda_{\mathrm{VB}}\right)e^{-},
	\label{eq:tetravalent_oxidation}
\end{equation} 
and the chemical etching of SiO$_2$ via
	\begin{equation}
	\mathrm{SiO_{2}} + \mathrm{6HF} \rightarrow \mathrm{SiF_{6}^{2-}}+2\mathrm{H_{2}O}+2\mathrm{H^{+}},
		\label{eq:chemical_etching}
	\end{equation} 
where $1\le \lambda_{\mathrm{VB}}\le 4$ is the amount of valence-band holes $h^{+}$ \cite{Zhang2001}. 

The experiments were conducted with an n-doped (1-10 $\Omega$cm) Si (111) sample as the working electrode in a three-electrode setup.  
The electrolyte was an aqueous solution containing 0.06 M NH$_4$F and 142 mM H$_2$SO$_4$. 
The uniformity of the electrode surface was monitored in-situ with an ellipsometric imaging setup which probes the change in optical path length at the electrochemical interface.
The changes in optical path length are converted into an intensity signal $\xi \left(\mathbf{x},t\right)$ ($\mathbf{x}$ denoting space and $t$ time) and recorded with a CCD-camera (JAI CV-A50).
We will present the ellipsometric intensity $\xi \left(\mathbf{x},t\right)$  as a percentage of the saturation threshold of the CCD camera. 
To allow for the oxidation of n-type Si, the electrode was illuminated with a linearly polarized He-Ne laser (HNL150L-EC, Thorlabs), the intensity of which was adjusted with a linear polarization filter. 
Further experimental details can be found in in the appendix~\ref{sec:Exp} and in Ref.~\cite{Patzauer2017}.

\begin{figure*}[htbp!]
    \includegraphics[width=0.95\textwidth]{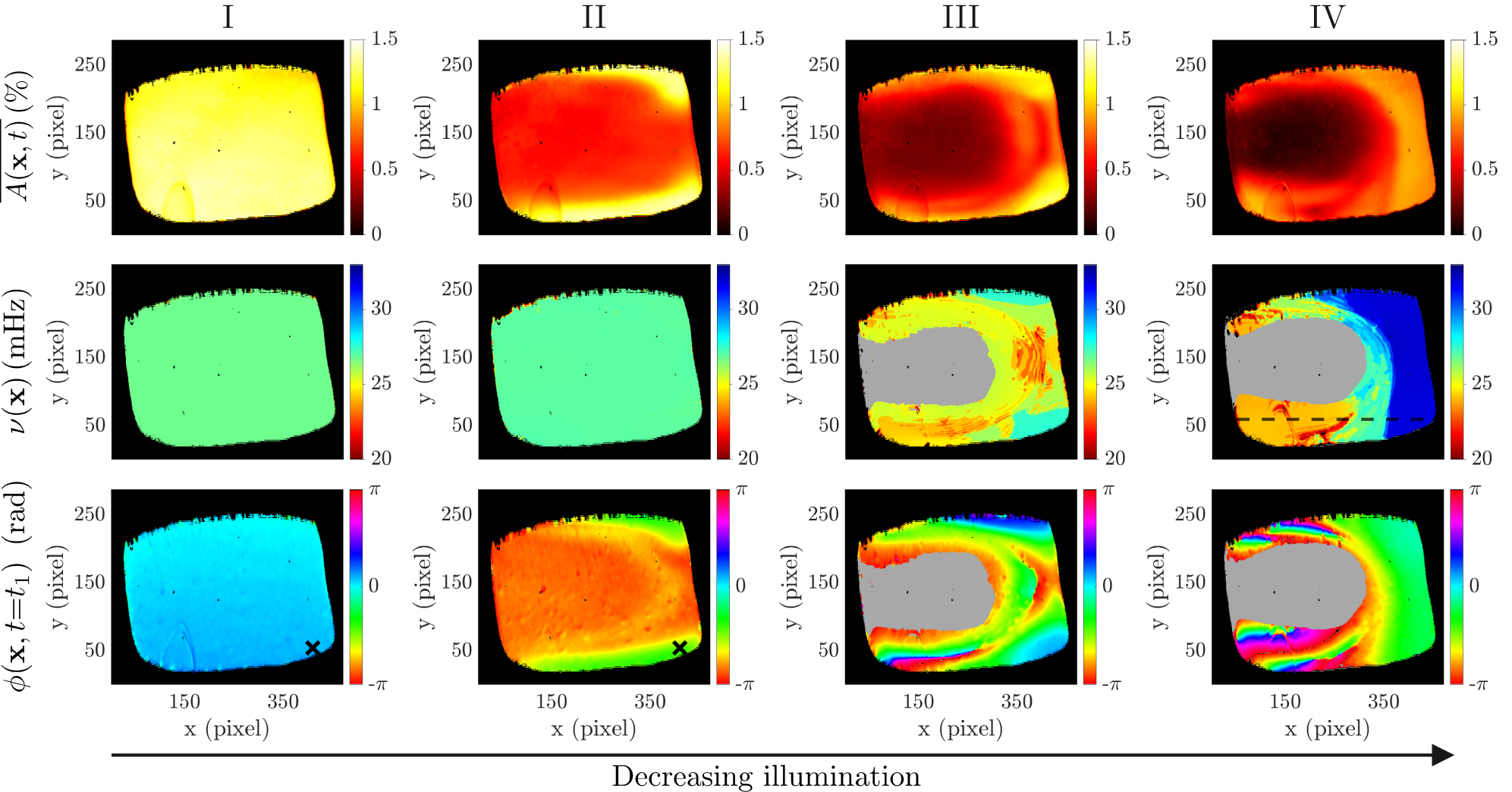}
    \caption{Experimentally measured  spatial distribution of the temporal average of the amplitude, $\overline{A\left(\mathbf{x},t\right)}$, of the dominant frequency $\nu\left(\mathbf{x}\right)$, and of the phase $\phi\left(\mathbf{x},t = t_{1}\right)$ at $t_{1} = 807$~s   at four different illumination intensities: I: $\mu = 5.97$~mW/cm$^{2}$, II: $\mu =0.95$~mW/cm$^{2}$, III: $\mu =0.80$~mW/cm$^{2}$, and IV: $\mu =0.73$~mW/cm$^{2}$. Electrode size: $A = 12.19$\,mm$^2$. \label{fig:nu_amp_phi}}
\end{figure*}

In the measurement presented below, the illumination intensity is the bifurcation parameter.
We initialized the electrode by applying a constant voltage at a high illumination intensity and then decreased the illumination step by step.
At each step, we waited until transients had died out and then measured the dynamics for $10^{3}$\,s. 
In order to characterize the dynamics of our system, we define the amplitude $A\left(\mathbf{x},t\right)$ and phase $\phi\left(\mathbf{x},t\right)$ of the ellipsometric intensity signal $\xi\left(\mathbf{x},t\right)$ at each pixel by calculating the analytic signal $\zeta \left(\mathbf{x},t\right)$ via the Hilbert transform $H\left(\xi\left(\mathbf{x},t\right)\right)$ (for details see Ref.~\cite{Pikovsky2001}):
\begin{equation}
\zeta \left(\mathbf{x},t\right) = \xi\left(\mathbf{x},t\right)+i H\left(\xi\left(\mathbf{x},t\right)\right) = A\left(\mathbf{x},t\right)\,e^{i\phi\left(\mathbf{x},t\right)}.
\label{eq:analy}
\end{equation}
Having determined the time-series of the amplitude and of the phase, we extracted the dominant frequency $\nu\left(\mathbf{x}\right)$ at each point from a linear fit to $\phi$ vs. t.

Exemplary states from a measurement series can be seen in Fig.~\ref{fig:nu_amp_phi} where the temporally averaged amplitude $\overline{A\left(\mathbf{x},t\right)}$, the dominant frequency $\nu\left(\mathbf{x}\right)$, and a snapshot of the phase at an arbitrary instant in time, $\phi\left(\mathbf{x}, t=807 \text{\,s}\right)$, are shown in the first, second, and third row respectively.
The four columns depict measurements at four different illumination intensities. 
The initial, highly illuminated state is shown in column I.
Here, the system oscillates uniformly with the same amplitude, frequency, and phase at each point in space, cf. Ref.~\cite{Schoenleber2014,Schoenleber2016}.

Upon lowering the illumination intensity (Fig.~\ref{fig:nu_amp_phi}~column II), the electrode splits into a region with higher and a region with lower amplitude. 
These two regions still oscillate with the same average frequency, but the oscillation phase differs between points in the higher- and lower-amplitude regions. 
In other words, amplitude clusters have formed.

\begin{figure}[tp!]
    \includegraphics[width=0.47\textwidth]{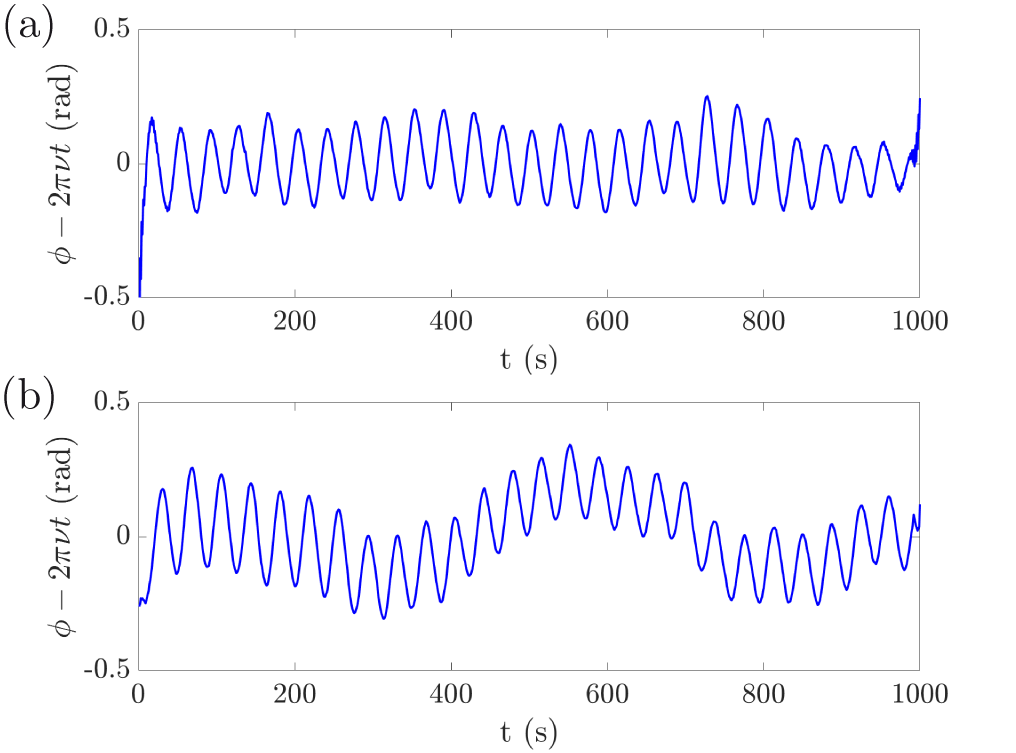}
    \caption{Time series of the local phase in a frame rotating with the dominant frequency $\nu$. 
    Time series taken at the point marked by a cross in the $\phi$ snapshot in Fig. \ref{fig:nu_amp_phi}\label{fig:MFR_phases} column~I and II respectively.}
\end{figure}

However, the data shown in Fig.~\ref{fig:nu_amp_phi}~column~II do not give the full picture of the dynamics. 
This can be seen if we look at Fig.~\ref{fig:MFR_phases}.
Here, the temporal evolution of the phase at an exemplary point, marked by a cross in the phase plots in Fig.~\ref{fig:nu_amp_phi}~column~I~and~II, is depicted in a frame rotating uniformly with the dominant frequency of the point in question.
Starting with the higher illumination (Fig.~\ref{fig:MFR_phases}(a)) we observe only a simple modulation with the same dominant frequency as the one of the rotating frame.
This is in fact the second harmonic of the dominant frequency of the original time series $\phi\left(t\right)$ and thus stems from its slight relaxational character.
In the case with the lower illumination (Fig.~\ref{fig:MFR_phases}(b)), when the amplitude clusters have formed, we observe a further slow modulation of the phase evolution. 
This suggests that the system not only underwent a pitchfork bifurcation leading to amplitude clusters but also a secondary Hopf bifurcation creating the modulated oscillations.

When lowering the illumination further, two drastic changes are observed (Fig.~\ref{fig:nu_amp_phi}~column~III). 
First, the mean amplitude differentiates further in space, suppressing the oscillations nearly completely on a part of the electrode. 
In this region, the very small amplitude combined with  experimental noise leads to apparent discontinuities in the phase, rendering the determination of the dominant frequency impossible. 
Therefore, in the second and the third row of Fig.~\ref{fig:nu_amp_phi}, we depict points with $\overline{A\left(\mathbf{x},t\right)}~<~0.35 \% $, in grey.
Second, and perhaps even more astonishing, focusing our attention on the region that exhibits well defined oscillations, $\overline{A\left(\mathbf{x},t\right)}~>~0.35 \% $, we observe that the dominant frequency is not uniform anymore. 
Rather, the frequencies appear to accumulate around three plateau values, as apparent from the turquoise, red, and yellow patches in Fig.~\ref{fig:nu_amp_phi}~column~III, whereby the higher frequencies are found in the regions with higher mean amplitude. 

In the last state (Fig.~\ref{fig:nu_amp_phi}~column~IV) the features that appeared in column III become more pronounced; on a part of the electrode the amplitude is practically completely suppressed.
In other words, on this part of the electrode we observe amplitude death \cite{Koseska2013}. 
Likewise, the frequency differences across the oscillating part of the electrode become more pronounced.
Equal, or at least very similar frequencies now appear in connected regions, whereby the frequency distributions of the two outer orange and blue regions are very narrow, and the frequency distribution of the middle, 'mediating region' is somewhat broader, ranging from light-blue to yellow. 
Indeed, we witness the self-organized formation of multi-frequency clusters in a homogeneous oscillatory medium. 
Considering the snapshot of the phase distribution, we observe that the faster region on the right oscillates nearly uniformly whereas the more slowly oscillating region on the left exhibits a continuous distribution of the phases over $2 \pi$\,rad. 
This travelling-wave-type feature can be seen as the continuum version of a splay state in networks of coupled oscillators.
Interestingly, the existence of mixed-type multi-frequency clusters consisting of a splay-type cluster and a phase-synchronized cluster, as we observe it here, has also been found in simulations of networks of phase oscillators with adaptive coupling, yet with the difference that the simulated phase-synchronized clusters occurred in antipodal pairs \cite{Kasatkin2017,Berner2018,Berner2019}.

A key to understanding the changes in the dynamics is to realize that our bifurcation parameter controls the effective number of degrees of freedom in the system.
At high illumination intensity $\mu$ there are more than sufficient valence-band holes for the oxidation process to take place equally everywhere on the electrode surface. 
Hence, the concentration of holes $n_h$ is effectively constant, and does not impact the uniform oscillation. 
The oscillations are synchronized by a global coupling arising from the presence of the external resistor and the potentiostatic control: 
\begin{equation}
\begin{aligned}
    \varphi^{\text{SC}}&\left(\mathbf{x},t\right) +  \varphi^{\text{OX}}\left(\mathbf{x},t\right)  = \\
    &=U- R \int\limits_{\mathbf{x}'\in A} i\left(\varphi^{\text{SC}}\left(\mathbf{x}',t\right),\varphi^{\text{OX}}\left(\mathbf{x}',t\right)\right)\,d\mathbf{x}'.
    \label{eq:URI}
\end{aligned}
\end{equation}
Here, $\varphi^{\text{SC}}\left(\mathbf{x},t\right)$ and $\varphi^{\text{OX}}\left(\mathbf{x},t\right)$ are the potential drops across the space charge layer of Si, and the SiO$_{2}$ oxide layer respectively, $U$ is the applied voltage, $R$ is the external resistance, $A$ is the electrode area, and $i$ is the local current.
The last term in Eq.~\eqref{eq:URI} describes the potential drop across the external resistor which depends on the total current.  
Since at high illuminations $\varphi^{\text{SC}}\left(\mathbf{x},t\right)$ remains constant, the oscillating total current causes oscillations in $\varphi^{\text{OX}}\left(\mathbf{x},t\right)$, which in turn influence the reaction rate and thus the oscillations. 
Hence, our electrochemical oscillator creates a mean field $I=\int\limits_{\mathbf{x}'\in A} i\left(t\right) \,d\mathbf{x}'$ that feeds back into the dynamics of the oscillations,
\begin{equation}
\dot{\xi}\left(\mathbf{x}, t\right) = F\left(\xi\left(\mathbf{x}, t\right), n_h, I; \mu\right).
\label{eq:meanfield}
\end{equation}
However, as we lower the illumination intensity $\mu$, $n_h$ becomes so low that, at some point, it starts to limit the reaction current. 
To compensate for the lower reaction rate, $\varphi^{\text{SC}}\left(\mathbf{x},t\right)$ increases and thus becomes time dependent as well. 
Hence, we face a situation where a physical quantity, namely $n_h$, starts to change in time  when the value of a parameter crosses a threshold: 
\begin{equation}
\dot{n}_h = G\left(n_h, I; \mu\right).
\label{eq:nu_dot}
\end{equation}
The dynamics of $n_h$ now depend on the oscillating mean field $I$ and also feeds back to said mean field as well as to the dynamics of the original 'base' oscillator.
Our oscillating medium is thus nonlinearly coupled as soon as $n_h$ becomes a degree of freedom of the dynamics.

If we consider our spatially continuous system as being composed of infinitesimally small base oscillators $\mathbf{w}_{k}$, one realizes that the nonlinear coupling is of the same type as the general physical setting for nonlinearly coupled oscillators formulated by Rosenblum and Pikovsky \cite{Rosenblum2007}: 
\begin{align}
    \dot{\mathbf{w}}_k &= \mathbf{F}\left(\mathbf{w}_{k},\mathbf{v},\mathbf{g};\mu\right), \label{eq:base_osc}\\
    \dot{\mathbf{v}} &= \mathbf{G}\left( \mathbf{v},\mathbf{g};\mu \right) \label{eq:coupling_dyn}.
    \end{align}
Here, $\mathbf{w}_{k}$ forms a base oscillator, the ensemble of all oscillators produces some mean fields $\mathbf{g}$, and $\mathbf{v}$ is a coupling variable that modulates the global coupling in a nonlinear way; $\mu$ is a bifurcation parameter. 
In our system, $\mu$ is the illumination intensity, $\mathbf{g}$ is the total current $I$, and $\mathbf{v}$ can be identified with $n_h$.

Nonetheless, in contrast to the global feedback variable $\mathbf{v}$ in Eq.~\eqref{eq:coupling_dyn}, $n_h$ is not strictly global, but rather a nonlocal variable which depends on space \cite{Patzauer2017}. 
Since $n_h$ influences the coupling strength of the base oscillator, cf. Eq.~\eqref{eq:meanfield}, the coupling becomes not only nonlinear but also space dependent. 
In this view, it is similar to the adaptive coupling discussed in Ref.~\cite{Kasatkin2017,Berner2018,Berner2019}, where multi-frequency clusters were found. 
We thus attribute the occurrence of our multi-frequency clusters to the combination of the nonlinear and nonlocal coupling, which allows for a self-organized adaptation of the coupling strength:
At parameter values at which multi-frequency clusters are found, the intra-cluster coupling strengths as well as all mutual inter-cluster coupling strengths differ.

\begin{figure}[tp!]
    \includegraphics[width=0.47\textwidth]{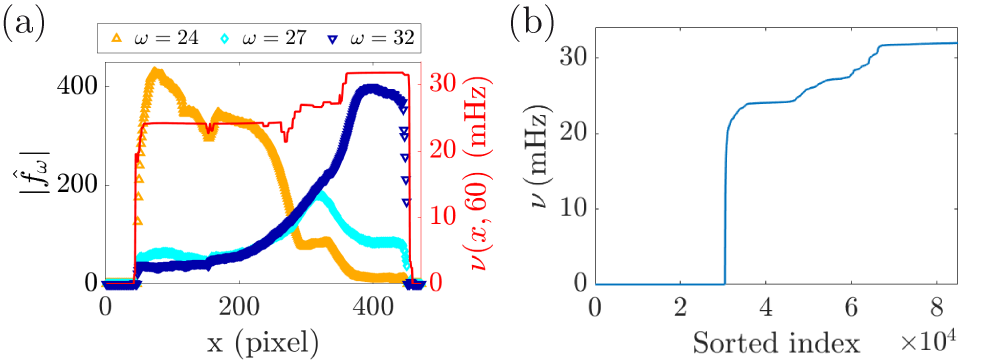}
    \caption{(a) Spatial profile of the absolute value of the Fourier transform of the ellipsometric intensity signal $\xi$ for three different frequencies (see legend) together with the dominant frequency $\nu$ (red). Profile taken at $y = 60 $, see dashed line in the $\nu$ plate in Fig.~\ref{fig:nu_amp_phi},~column~IV. (b) The dominant frequencies $\nu $ from Fig.~\ref{fig:nu_amp_phi} IV sorted in ascending order.
    \label{fig:fourier_and_sorted}}
\end{figure}

Adopting a different perspective, one realizes that our dynamics also contain features that have been discussed in connection with certain types of chimera states. 
Provata considered a birhythmic model \cite{Provata2020}.
When coupling these oscillators nonlocally, synchronized domains oscillating in either of the two bistable limit cycles could be stabilized. 
The interfacial regions mediating between the domains with different frequencies oscillated asynchronously with frequency components of both adjacent regions. 
Provata interpreted her two frequency-domains separated by a 'more frequency' incoherent region as a chimera state.
Our multi-frequency cluster (Fig. \ref{fig:nu_amp_phi}, column IV) exhibits the same features. 
This can be seen in Fig.~\ref{fig:fourier_and_sorted}(a) where we present the absolute value of the Fourier coefficients of the three main frequencies, 24\,mHz, 27\,mHz, and 32\,mHz, of the local Fourier spectra along a the dashed line in the $\nu$ plot in Fig.~\ref{fig:nu_amp_phi}~collumn~IV.
While in the left low- and the right high-frequency regions the contribution of the other two frequencies are very small, in the middle region we find not only the third, dominant frequency at 27\,mHz, but also a significant contribution of the frequencies of the two adjacent regions, just as in Provata's model system.
In our case, the dynamics of the mediating region appears phenomenologically rather coherent. 
On this basis, we would not classify our state as a typical chimera state, in contrast to the chimera states observed previously during silicon photo-electrodissolution \cite{Schoenleber2014,Schmidt2014}.

However, the classification of chimera states is a very intricate matter.
This becomes clear when considering the relation of multi-frequency clusters of different spatial extensions in continuous systems to weak chimera states in ensembles of four coupled discrete oscillators \cite{Ashwin2015}. 
Loosely speaking, a chimera state in such a system is characterized by two oscillators being synchronized, with the frequency $\nu_1$, while the other two oscillators possess frequencies which differ from each other as well as from $\nu_1$ \cite{Ashwin2015}. 
If, to a first approximation, we neglect the interfacial regions, a multi-frequency cluster state with three different regions in a spatially extended system can likely be reduced to a low-dimensional system of three coupled oscillators where the coupling is weighted by the size of the domains.
Thus, the dynamics of multi-frequency clusters in systems with many degrees of freedom is in a sense equivalent to the one of a weak chimera state. 
With this in mind it appears worthwhile 
to differentiate between chimera states where the number of incoherent oscillators scales with the system size and states where it does not.
Kemeth et al. have presented considerations along these lines and coined the first type of chimera state \textit{extensive} which would suggest that the three frequency cluster state could be classified as an \textit{intensive} chimera state \cite{Kemeth2018}.
In this respect, an important question to be investigated in the future is whether one can identify general dynamical properties that determine whether a weak chimera state behaves as an intensive or extensive chimera, in the sense defined here when successively increasing the number of oscillators. 

Another issue concerning the classification of the dynamics arises when regarding only the global picture of the dominant frequencies, neglecting any spatial information, as shown in Fig.~\ref{fig:fourier_and_sorted}(b). 
Here, the dominant frequencies as found in state IV of Fig.~\ref{fig:nu_amp_phi} are sorted in ascending order.  
The first about thirty thousand entries with the value $0$~Hz arise from the region where we observe amplitude death. 
For higher indices, we clearly observe three plateaus.
These reflect our three frequency domains. 
However, the transitions between these plateaus are not sharp but instead occur continuously in a finite index range. 
As such, this graph is reminiscent of the distribution of dominant frequencies in 2-frequency chimera states \cite{Dai2017,Zakharova2019,Suda2020}, and to some extent also of the ones in hybrid chimera states \cite{Bolotov2018}, which are composed of a fully synchronized, a nearly-synchronized, and an incoherent part. 
Hence, Fig.~\ref{fig:fourier_and_sorted}(b) shows again that there might be aspects in the dynamics of multi-frequency clusters in continuous media that are related to those of chimera states. 

In conclusion, our experimental observation of multi-frequency clusters is not only an exceptional example where a self-organized adaptive coupling was observed in a non-living system, but also reveals important open problems concerning the properties of multi-frequency states in continuous systems, such as their relation to chimera states or requirements on the adaptive coupling for their existence.

\begin{acknowledgments}
The authors would like to thank Felix P. Kemeth, Seungjae Lee, Munir M. Salman, Anton Tosolini, Sindre W. Haugland, and Juliane Wiehl for fruitful discussions.
This work has been supported by the Deutsche Forschungsgemeinschaft project KR1189/18 "Chimera States and Beyond".
\end{acknowledgments}
\bibliography{f-paper-bibliography}
\appendix
\section{\label{sec:Exp}Experimental}
We used a three-electrode setup with an n-doped \mbox{(1-10~$\Omega$cm)} single crystalline (111) Si sample as the working electrode.
Before the experiments, the Si sample was equipped with a back contact by thermally evaporating aluminum on to the back and then annealing it at 250$^{\circ}$\,C for 15\,min.
Then the front side of the electrode was treated with an oxygen plasma in order to rid it of any organic contamination.
The sample was then mounted on a custom-made PTFE sample holder using a conductive silver paste and sealed using red silicone rubber (Scrintex 901, Ralicks GmbH);
15-25\,mm$^2$ of the Si sample were left exposed, forming the active area of the working electrode.
The active area was cleaned by  wiping the electrode with acetone-drenched precision wipes and sequentially immersing it in acetone, ethanol, methanol, and ultrapure water (18.2 M$\Omega$cm) for 10\,min each.

The mounted electrode was then placed in the center of the cell with the Hg$|$Hg$_2$SO$_4$ reference electrode placed behind it.
For the counter electrode, we bent a Pt wire (99.99\,\% Chempur) into a circle and placed it symmetrically in front of the working electrode.
In order to control the voltage between working and reference electrode, we used a FHI-2740 potentiostat (electronics laboratory of the Fritz-Haber-Institut, Berlin, Germany).

The aqueous electrolyte had a total volume of 500\,ml and contained 0.06\,M NH$_4$F and 142\,mM H$_2$SO$_4$, yielding a pH of 1 in accordance with the dissociation constants found in Ref. \cite{Cattarin2000}.
The electrolyte was purged with argon for 30\,min before the experiment and an argon overpressure was kept throughout all measurements via a gas inlet above the electrolyte.
The electrolyte was also stirred using a magnetic stirrer at 20\,Hz throughout all measurements. 

All glassware was cleaned in HNO$_3$ and subsequently in a 1\,M aqueous KOH solution and stored in ultrapure water.
Platinum and PTFE parts were cleaned in Piranha solution.
All organic cleaning solvents were AnalaR NORMAPUR grade (VWR Chemicals). 
All electrolyte components were Suprapur grade (Merck).
\subsection{Ellipsometric imaging}
\begin{figure}[!tp]
	\includegraphics[width=0.8\columnwidth]{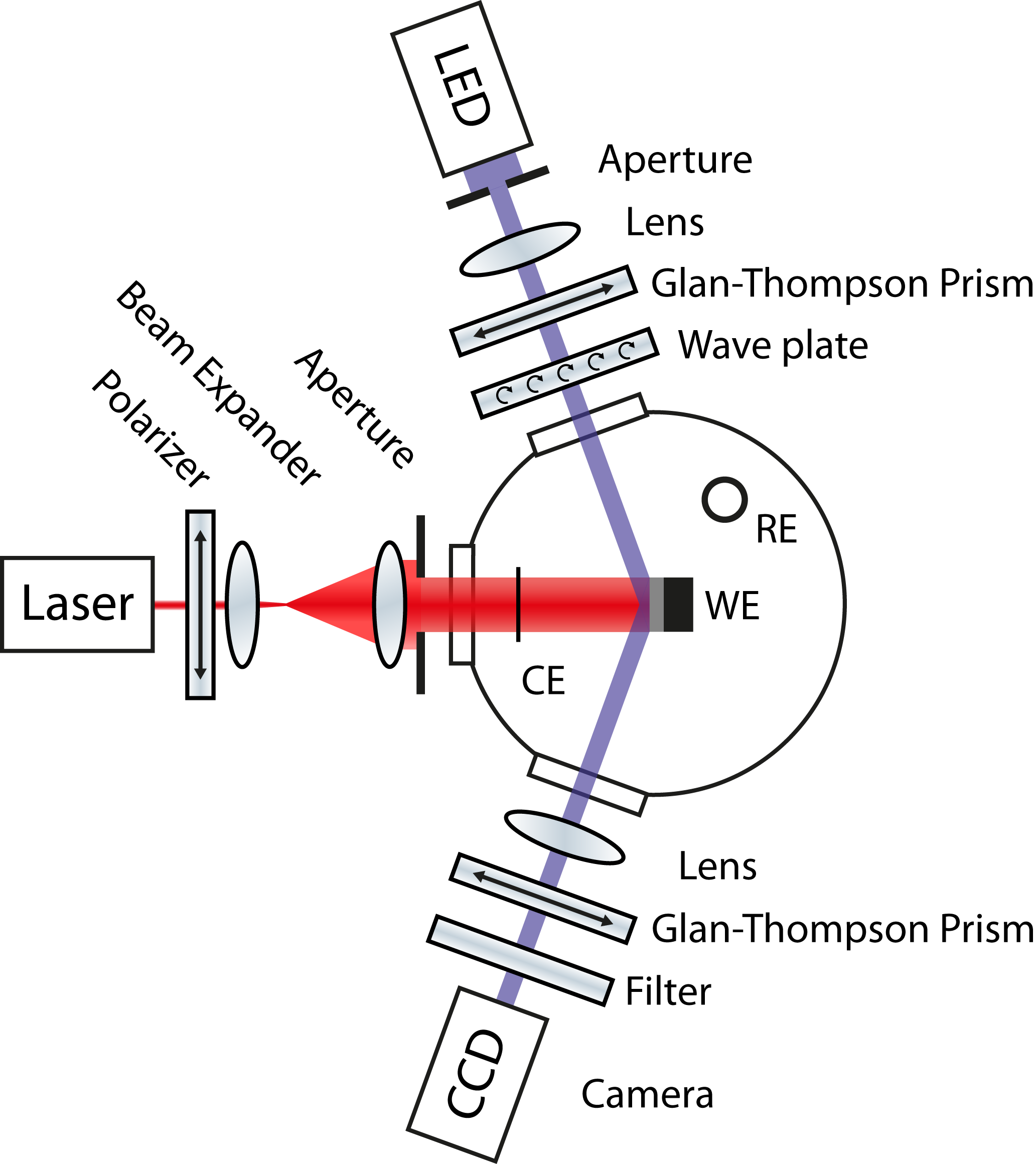}
	\caption{Sketch of the three electrode electrochemical cell showing the position of the working electrode (WE), the reference electrode (RE), and the counter electrode (CE) as well as both the ellipsometric imaging setup and the laser illumination setup. The light from the LED (blue) is first elliptically polarized and then reflected off the WE, resulting in a change in polarization, depending on the optical path length through the oxide. The change in polarization is then converted into an intensity signal which is measured by the CCD camera. The intensity of the laser-light used for illumination of the WE is controlled via the polarizer next to the laser.\label{fig:EMSI_setup}}
\end{figure}

We used the ellipsometric imaging setup sketched in Fig.~\ref{fig:EMSI_setup} to monitor changes in the optical path through the oxide layer in-situ.
The non-polarized light coming from the LED (Linos, HiLED, 470\,nm) becomes elliptically polarized once it passes through the Glan-Thompson prism and the zero-order $\lambda$/4 plate.
The beam is then reflected off the working electrode at an angle close to the Brewster angle of water and Si $\alpha=70^{\circ}$.
Depending on the length of the optical path at the electrochemical interface, the ratio between the s- and the p- polarized components of the light change.
The polarization is then converted into an intensity signal by letting the light pass through a second Glan-Thompson filter.
The intensity was measured using a CCD-camera (JAI CV-A50) and digitized using a frame grabber card (PCI-1405, National Instruments).
The spatial average of the frame was sampled at 10\,Hz and one spatially resolved frame was saved each second.
The CCD gives a linear response to the intensity of the incoming illumination, up to a saturation threshold; we present the ellipsometric intensity as a percentage of this threshold.

In general, the light intensity from the LED varies slightly across the electrode. 
This leads to a variation of the raw ellipsometric intensity $\xi \left(\mathbf{x},t\right)_{\textnormal{raw}}$ depending on the position $\mathbf{x}$ on the electrode.
To adjust for this variation, a background correction was applied by subtracting the temporal average of the raw data $\overline{\xi \left(\mathbf{x}\right)}_{\textnormal{raw}}$ individually at every point.
In addition to this background variation of the intensity, the contrast positively correlates with the absolute value of the LED illumination intensity. 
Hence, a point on the sample that is illuminated with a high background intensity will have a higher contrast. 
To counter this, we correct each individual pixel by dividing its value by its temporal average. This correction factor is then normalized by multiplying with the spatial average of the temporal average of the entire image.

In total, the correction suppresses the signal from pixels with high temporal average and enhances the signal from pixels with low temporal average.
The complete background correction is summarized in Eq.~\eqref{eq:raw_data_sup}:
\begin{equation}
    \xi \left(\mathbf{x},t\right) = 
    \left(\xi \left(\mathbf{x},t\right)_{\textnormal{raw}}-
    \overline{\xi \left(\mathbf{x},t\right)}_{\textnormal{raw}}\right) \frac{\langle\overline{\xi\left(\mathbf{x},t\right)}_{\textnormal{raw}}\rangle}{\overline{\xi \left(\mathbf{x},t\right)}_{\textnormal{raw}}},
\label{eq:raw_data_sup}
\end{equation}
with $\xi \left(\mathbf{x},t\right)$ being the corrected local time series and $\langle\overline{\xi\left(\mathbf{x},t\right)}_{\textnormal{raw}}\rangle $ the spatial average of the temporal average of the raw data.
To reduce the noise, we smoothed the data in the temporal direction by using a Savitzky-Golay filter with a 2nd degree polynomial and a 15 point window.
In addition, the data was binned into 5x5 pixels bins.
\subsection{Illumination}
Since n-type Si mainly interacts with the electrolyte through valence-band processes (see Eq.~\eqref{eq:tetravalent_oxidation} in the manuscript), the sample had to be illuminated to allow for anodic oxidation.
For this purpose, a linearly polarized He-Ne laser (HNL150L-EC, Thorlabs) was used.
The fact that the laser was linearly polarized allowed us to adjust the illumination intensity with a polarization filter mounted on a motorized rotation mount (KPRM1E/M, Thorlabs) and placed directly after the laser.
After the polarizer, the beam was widened using a beam expander and then passed through an iris diaphragm.
This allowed for the illumination of the entire sample with the central, more uniform, part of the beam.
A sketch of the illumination setup can be seen in Fig.~\ref{fig:EMSI_setup}.
Note that, the intensity of the laser was much higher than the intensity of the LED used for the ellipsometric imaging.

At the beginning of the measurement series, we initialized the electrode in a uniformly oscillating state via a potential step from OCP to $U_{\text{app}}=4.15$\,V~vs~SHE, and then held the voltage constant whilst illuminating with a high illumination intensity.

\end{document}